\documentstyle[12pt,fleqn]{article}
\begin{document}
\baselineskip= 22 truept
\def\be{\begin{equation}}
\def\ee{\end{equation}}
\def\bea{\begin{eqnarray}}
\def\eea{\end{eqnarray}}
\def\pa{\partial}
\def\dmu{\partial_\mu}
\def\A{\tilde{A}}
\def\k{{\cal K}}
\def\h{{\cal H}}
\def\p{{\cal P}}
\def\q{{\cal Q}}
\def\n{{\cal N}}
\def\rt{\rightarrow}
\begin{titlepage}
\begin{flushright}
IP/BBSR/95-51 \\
hep-th/9506037 \\
\end{flushright}
\vspace{1cm}\begin{center} {\large \bf Symmetries of
Heterotic String Theory}\\ \vspace{1cm} {\bf Anindya K.
Biswas${}^1$, Alok Kumar${}^{1,2}$ and Koushik Ray${}^1$ }
\\ 1. Institute of Physics,{\footnote  {Permanent address}}
\\Bhubaneswar 751 005, INDIA \\ email: akb, kumar,
koushik@iopb.ernet.in \\
2. ICTP, 34100 Trieste, Italy \\
email: alok@ictp.trieste.it \\
\today
\end{center}
\thispagestyle{empty}
\vskip 4cm
\begin{abstract}
We study the symmetries of the two dimensional
Heterotic string theory by following the approach of
Kinnersley et al for the study of
stationary axially-symmetric Einstein-Maxwell equations.
We identify the finite dimensional symmetries which
are the analogues of the groups $G'$ and $H'$ for
the Einstein-Maxwell equations. We also give the
constructions for the infinite number of conserved currents
and the affine $\hat{o}(8, 24)$ symmetry algebra
in this formulation. The generalized Ehlers
and Harrison transformations are identified and a
parallel between the infinite dimensional symmetry algebra
for the heterotic string case with $\hat{sl}(3, R)$ that arise
in the case of Einstein-Maxwell equations is
pointed out.
\end{abstract}
\vfil
\end{titlepage}
\eject
\section{Introduction}
There has been a resurgence of interest in
the investigatons of duality symmetries \cite{sch1,wit,nwit}
in string theory.
In a series of interesting papers, Sen gave  the explicit
realization of S- and T-dualities for the
torus compactified heterotic string
theories in various dimensions\cite{sen4,sch,sen3,sen2,sen6},
exploited them for obtaining the solitonic spectrum of
these theories and argued in favour of the nonperturbative
validity of the S-duality symmetry.
Partly inspired by these results,
further studies led to the interesting
developments towards understanding the duality symmetries
of other supersymmetric gauge and string theories
\cite{seib}.
It is now believed that the symmetries like S-duality are
the symmetries of more genral string theories and
can possibly help in handling their
nonperturbative aspects. More recently, duality
symmetries relating type II and heterotic string theories
\cite{wit,sen6} as
well as the ones combining S- and T-duality into a bigger
symmetry, namely U-duality\cite{hull},
have also been discovered. The
enlarged duality symmetries are also known to be present in
dimensions other than four\cite{hull,wit,sen3,sen2}.
For example, in three dimensions,
the T-duality $O(7, 23; Z)$ combines with the S-duality
to form a bigger group $O(8, 24; Z)$\cite{sen3}.
In this paper, we study the duality symmetries of the two
dimensional heterotic string theory by exploiting
their resemblance to the Einstein-Maxwell theory
with two commuting isometries.

Symmetries of the stationary axisymmetric Einstein-Maxwell
equations have been studied a great deal
over the years in the past\cite{xan,ernst}. After the
pioneering work by Geroch\cite{geroch} for the stationary, axisymmetric
Einstein equations, in a series of papers,
Kinnersley et al\cite{kinn1,kinn2,kinn3,kinn4}
unravelled a very rich symmetry structure in the
stationary, axisymmetric Einstein-Maxwell equations, or
equivalently, the so-called Ernst equations --- a formulation due
to Ernst. In fact, they showed that these equations enjoy an
infinite-dimensional symmetry and found  that the
algebra is an infinite extension of the $sl(3, R)$
algebra. It was realized later that this was the
current algebra $\hat{sl}(3, R)$ and it is this symmetry that
renders the non-linear Ernst equations integrable. There exists
an infinite number of conserved currrents in this theory and the
algebra is realized in terms of them. A parallel of these
studies has also been carried out for supergravity theories
compactified to two dimensions.\cite{hn,nicolai}

Some aspects of the Ernst sigma model and the infinite
dimensional symmetries have been analyzed recently in the
context of heterotic string theory\cite{bakas,pui,jmu,sen2}.
In particular, the T- and S-duality
symmetries of the four dimensional
string theory in the presence of two commuting isometries
have been shown to be a subgroup of the infinite
dimensional String Geroch group\cite{bakas,sen2}.
It was also shown in \cite{sen2} that the $O(8,
24)$ symmetry of the heterotic string theory in 3-dimensions
\cite{sen3} has an affinization to $\hat{O}(8,24)$ in
2-dimensions.
However in these extensions to string theories, the
emphasis has mainly been on the algebraic
and symmetry structure  and comparatively
less on their applications to the classification and
the generation of the background fields. More recently some
work has been carried out to use these transformations
for practical purposes
in theories with axion and dilaton fields\cite{galtsov}.
The full structure of
the heterotic string theory, however, has been missing in these
attempts since the moduli fields are not included.

In the approach by Kinnersley et,
al\cite{kinn1,kinn2,kinn3,kinn4} to
the investigation of the Einstein-Maxwell therory with two
commuting isometries, both the algebraic
structure of the symmetries as well as their applications
to the gravity backgrounds are clarified. For example,
the finite form of the symmetry transformations which can
be used for generating solutions have been given in
\cite{kinn1,kinn3}. They include interesting nontrivial
transformations such as the Ehlers and the Harrison transformatins.
The infinitesimal form of these transformations
on an infinite set of potentials responsible for
generating the affine algebra is also given in these papers.

In a previous paper\cite{ray},
two of the present authors applied the
formalism of Kinnersley et al to the
studies of the string effective action and found a close analogy
between the equations of motion of the
Einstein gravity and the String effective
action for vanishing $E_8\times E_8$ backgrounds.
Based on this analogy, we gave the analogues of the
Ehlers transformation for the string effective action and
showed their connection to the $O(d, d)$ symmetry
transformations. We in fact demonstrated that the
generalized Ehlers transformation
of string theory can be identified
with the coset ${O(d)\times O(d)}\over {O(d)}$ of the
symmetry group $O(d, d)$.  This $O(d, d)$
, unlike T-duality, however acts on
the variables $ \psi$, $G$ and $\phi$, where $\psi$ is the
field
dual to the the antisymmetric moduli $B$. $G$ are the symmetric
moduli and $\phi$ is the dilaton field. The action of the
Ehlers transformation on $B$ is given by its
induced action through a duality relation:
\begin{equation}
\rho G^{-1} \partial_\mu B G^{-1} = - \epsilon_{\mu \nu }
\partial^\nu \psi. \\ \label{si2}
\end{equation}
It was found that the $O(d, d)$ transformations can then
be classified into three sets of $O(d, d)$ matrices satisying
\begin{equation}
\Omega^T L \Omega = L,	\quad
L = \left(\begin{array}{cc} 0 & I_d \\
			I_d & 0
\end{array} \right)
\label{domega}
\end{equation}
as
\begin{equation}
\Omega_1 = \left(\begin{array}{cc} I_d & \gamma \\
                          0 & I_d \\
\end{array}\right)  \quad
\Omega_2 = \left(\begin{array}{cc} I_d & 0 \\
                          \alpha & I_d \\
\end{array}\right) \quad
\Omega_3 = \left(\begin{array}{cc}  {A^{-1}}^T & 0 \\
                                        0 & A \\
\end{array}\right),	\label{omega}
\end{equation}
where $I_d$ is the $d$-dimensional identity matrix.
Among these, $\Omega_2$ and $\Omega_3$ correspond
to the constant shift in the antisymmetric moduli and the
constant coordinate transformations respectively. $\Omega_1$
are the only transformations which act differently on
$B$ and on $\psi$. They act non-locally on $B$ and are
responsible for the presence of the infinite set of
conserved currents. These are identified as the
Ehlers transformation for string theory.

In this paper we extend the analysis of our previous work to
include nontrivial gauge fields of heterotic string
theory. We once again show a close analogy of the
equations of motion of heterotic string theory with
the analysis of \cite{kinn1,kinn2}. We generalize the
finite dimensional groups $H'$ and $G'$
in their analysis of
Einstein-Maxwell theory to the heterotic string
theory. These groups in string case act manifestly
on the original and the
dual equations of motion respectively.
We also explicitly
construct the infinite number of conserved currents
of this theory and their transformation rules
starting from original as well as dual equations of
motion. We identify the
infinite dimensional symmetry algebra of this theory with
the affine $\hat{o}(8, 24)$ algebra and identify the analogues of
the Ehlers and Harrison transformations. The duality symmetry
of the two dimensional heterotic string theory is the discrete
subgroup $\hat{O}(8, 24; Z)$ of this infinite dimensional
symmetry\cite{sen2}.

The rest of the paper is planned as follows. In section-2 we
give the equations of motion of the heterotic string and show their
analogy with that of the Einstein-Maxwell equations. We also
write down the equations of motions using the dual
variables and show that they are also analogous to the
equations in the Einstein-Maxwell theory. Section-3 gives the
construction of the infinite set of potentials. The symmetry
transformations, the related algebra and their identification
with $\hat{o}(8, 24)$ is given in section-4. These
correspond to the enlargements of the $G'$ and $H'$
to the infinite dimensional group $K'$ identified as
$\hat{O}(8, 24)$ in the present case. The conclusions and the
discussions are given in section-5.

\section{Derivation of the Equations}

We now begin our study of the bosonic part of the two dimensional
heterotic string theory with nontrivial dilaton and
moduli fields which include the nontrivial
$E_8\times E_8$ gauge field backgrounds of the
ten dimensional heterotic string theory compactified to
two dimensions. The situation is equivalent to
considering the four dimensional heterotic string theory
with two commuting isometries. Furthermore,
since the gauge fields have no dynamics in two dimensions, the
two dimensional gauge fields are chosen to have vanishing
backgrounds\cite{kinn1,sen2}. We shall start,
therefore, with the 2-dimensional effective action of the
heterotic string theory, namely,
\begin{equation} \label{action}
S = \int d^2 x \sqrt{g} e^{-2\phi} \left[ R + 4 g^{\mu \nu}
\partial_{\mu}\phi \partial_\nu\phi + \frac{1}{8} g^{\mu \nu}
Tr( \partial_\mu {\cal M}^{-1} \partial_\nu {\cal M}) \right],
\end{equation}
where the matrix ${\cal M}$, representing the moduli $G$, $B$ and
$A$ is parametrized as:
\begin{equation}
\label{defcalm}
{\cal M} = \left( \begin{array}{ccc} G^{-1} & G^{-1} (B + C) & G^{-1} A
\\ (-B + C) G ^{-1} & (G - B + C) G^{-1} (G + B + C) & (G - B +
C) G^{-1} A \\ A^T G^{-1} & A^T G^{-1} (G + B + C) & I_{16} + A^T
G^{-1} A  \end{array} \right).
\end{equation}
$G$ and $B$ are respectively $d\times d$ symmetric and
antisymmetric matrix-valued scalar fields. $A$ denotes
$d\times n$ matrix-valued scalar fields coming from
the gauge fields of the heterotic strings.
For the heterotic string theory,
$d = 8$ and $n=16$, but we keep these parameters
arbitrary in the present
discussion to avoid confusion among indices when they are used.
$C = \frac{1}{2} A A^T$ is a
$d\times d$ matrix.
The equations of motion for the moduli ${\cal M}$ and dilaton $e^{-2
\phi}$ derived from the above action can be written as
\begin{eqnarray}
\label{eom1}
\partial_{\mu} ( \sqrt{g} g^{\mu \nu }
e^{-2\phi} {\cal M}^{-1} \partial_\nu
{\cal M} ) = 0, \\ \label{eom2}
\partial_\mu ( \sqrt{g} g^{\mu \nu} \partial_\nu e^{-2\phi} ) =
0, \\
\label{eom3}
R^{(2)}_{\mu \nu} + 2\nabla_\mu \nabla_\nu \phi + \frac{1}{8} Tr
(\partial_\mu {\cal M}^{-1} \partial_\nu {\cal M}) = 0.
\end{eqnarray}
The equations (\ref{eom1}) and (\ref{eom2}) can be
rewritten in the conformal gauge, with an identification
$ \rho = e^{-2\phi}$, as
\begin{eqnarray}
\label{eom11}
\partial^{\mu} ( \rho {\cal M}^{-1} \partial_\mu {\cal M} ) = 0, \\
\label{eom22} \partial^\mu \partial_\mu \rho = 0.
\end{eqnarray}
${\cal M}$ also has the properties:
\be
{\cal M}^T = {\cal M}, \qquad
{\cal L}{\cal M}{\cal L} = {\cal M}^{-1},
\ee
where $L $ is given by
\begin{equation} \label{rest}
 {\cal L} = \left( \begin{array}{ccc} 0 & I_d & 0 \\
                               I_d & 0 & 0 \\
                               0 & 0 & I_{n} \\
\end{array}     \right),
\end{equation}
and $I_n$ denotes an $n \times n$ identity matrix.
The third equation (\ref{eom3}) turns out to be an integrable
one for the conformal factor and takes the following form in the
conformal gauge $ g_{\mu \nu } = e^{2\Gamma} \delta_{\mu \nu} $:
\begin{eqnarray} \label{eom33}
- \delta_{\mu \nu} \partial^\sigma \partial_\sigma \Gamma + 2
(\partial_\mu \partial_\nu \phi &-& \partial_\mu \phi \partial_\nu
\Gamma - \partial_\nu \phi \partial_\mu \Gamma + \delta_{\mu \nu}
\partial^\sigma \Gamma \partial_\sigma \phi) \nonumber \\
&+& \frac{1}{8} Tr ( \partial_\mu {\cal M}^{-1}
\partial_\nu {\cal M}) = 0.
\end{eqnarray}
Here we are choosing the two dimensional metric to be
Euclidean. This corresponds to the contraction to the
Levi-Civita tensor: $\epsilon^{\mu \alpha}
\epsilon_{\mu \beta} = \delta^{\alpha }_{\beta}$. All
the results, however, are valid in Minkowski space as
well with few changes of signs  in the formulas
presented here.
Determination of this conformal factor is important for the
generation of new solutions.
However, since this equation for $\Gamma$ is integrable,
we only concentrate on equations
(\ref{eom11}), (\ref{eom22}) for studying the symmetries
of the theory. These
equation are the same as that for the Ernst sigma
model\cite{xan,nicolai,bakas}. It is also clear that they
are invariant under an $O(d, d+n)$ group of symmetry
transformations:
\begin{equation}
\label{o824}
{\cal M} \rightarrow \tilde{{\cal M}} = \Omega {\cal M} \Omega^T,
\end{equation}
where $\Omega$ satisfies
\begin{equation}
\Omega^T {\cal L} \Omega = {\cal L},
\end{equation}
which includes the T-duality symmetry of the string theory as its
discrete subgroup $O(d, d+n;Z)$ and has been used for generating
solutions. The non-trivial action of T-duality is represented
by a group $O(d;Z) \times O(d+n;Z) \over O(d;Z)$\cite{asen}. The
T-duality is
also conjectured to be a symmetry, not only of the string
effective action, but also of the full string theory\cite{sch}.
However it  has already been pointed out\cite{sen2} that in two
dimensions, which is the case at hand, the symmetry is not
merely $O(8, 24) $, but an infinite dimensional one, viz
$\hat{O}(8, 24)$.
We shall show the existence of $\hat{O}(8, 24)$ symmetry
in later sections
after dualizing equations (\ref{eom11})--(\ref{eom22}).

At this point we would like to
mention to the readers interested in seeing only
the $\hat{O}(8, 24)$ symmetry structure that they can
directly proceed to the analysis
of sections (3.b) and (4.b). However the full structure
of the analysis by Kinnersley et al is unravelled only by
studying the component forms of equation (\ref{eom11}) and
their dualizations. This is pursued in the following.
The connection between the
symmetry structures in the original and the dual formulations
of supergravity also requires the full analysis of sections-3
and 4.
In section-5 we comment on the comparison between the
dualization procedure given below and the one in section (3.b).

The symmetry transformations on the moduli is now determined on the
lines of \cite{kinn1,kinn2} by obtaining the
twist potentials. For this we
observe that equations (\ref{eom11}) are total divergence
conditions and when written in terms of $G$, $B$, and $A$ take the
form,
\bea
\label{eqn1}
\pa^\mu \left[ \rho ( \dmu E^T G^{-1} + E^T G^{-1} (\dmu B + \dmu C -
\dmu A A^T ) G^{-1} ) \right] &=& 0, \\
\label{eqn2}
\pa^\mu \left[ \rho ( E^T G^{-1} \dmu E - \dmu E^T G^{-1} E - E^T
G^{-1} (\dmu B + \dmu C - \dmu A A^T ) G^{-1} E ) \right] & = & 0, \\
\label{eqn3}
\pa^\mu \left[ \rho ( E^T G^{-1} \dmu A - \dmu E^T G^{-1} A - E^T
G^{-1} (\dmu B + \dmu C - \dmu A A^T ) G^{-1} A ) \right] & = & 0,
\\ \label{eqn4}
\pa^\mu \left[ \rho ( G^{-1} \dmu E - G^{-1} (\dmu B + \dmu C - \dmu A
A^T ) G^{-1} E ) \right] & = & 0, \\
\label{eqn5}
\pa^\mu \left[ \rho ( G^{-1} \dmu A - G^{-1} (\dmu B + \dmu C - \dmu A
A^T ) G^{-1} A ) \right] & = & 0, \\
\label{eqn6}
\pa^\mu \left[ \rho ( A^T G^{-1} \dmu A - \dmu A^T G^{-1} A - A^T
G^{-1} (\dmu B + \dmu C - \dmu A A^T ) G^{-1} A ) \right] & = & 0,
\\  \label{eqn7}
\pa^\mu \left[ \rho ( G^{-1} (\dmu B + \dmu C - \dmu A A^T ) G^{-1}
) \right] & = & 0, \\ \label{eqn8}
\pa^\mu \left[ \rho ( \dmu A^T G^{-1} + A^T G^{-1} (\dmu B + \dmu C -
\dmu A A^T ) G^{-1} ) \right] & = & 0, \\
\label{eqn9}
\pa^\mu \left[ \rho ( A^T G^{-1} \dmu E - \dmu A^T G^{-1} E - A^T
G^{-1} (\dmu B + \dmu C - \dmu A A^T ) G^{-1} E ) \right] & = & 0,
\eea
where $E \equiv G + B + C$,
and superscript ${}^T$ denotes the transpose.
Note that these equations are not all independent. Equations
(\ref{eqn1}), (\ref{eqn3}) and (\ref{eqn5}) are the
transpose of
equations (\ref{eqn4}), (\ref{eqn9}) and (\ref{eqn8}),
respectively. This leaves us with six equations in three
unknowns, viz. $G$, $B$ and $A$. However, they are also not all
algebraically independent. For example, equation(\ref{eqn3}) is
identically satisfied using (\ref{eqn1}) and (\ref{eqn5}).
We shall take equations (\ref{eqn4}), (\ref{eqn5}) and
(\ref{eqn7}) as three independent equations.
The reduction to only three independent equations
is due to the fact that eqn.(\ref{eom1}) is
basically the
$O(d, d+n)$ invariant combination of the three equations of motion for
$G$, $B$ and $A$.
Equations (\ref{eqn4}) can be further simplified using the other
two, so that the independent set of equations for the
heterotic string theory can be written as:
\begin{eqnarray}
\label{feqn1}
\pa^{\mu}[\rho G^{-1}(\pa_{\mu}B - {1\over 2}\pa_{\mu}A A^T
	+ {1\over 2}A\pa_{\mu}A^T)G^{-1}] = 0, \\
\label{feqn2}
\pa^{\mu}(\rho G^{-1}\pa_{\mu}A)
	- \rho G^{-1} (\pa_{\mu}B - {1\over 2}\pa_{\mu}A A^T +
	{1\over 2}A\pa_{\mu}A^T)G^{-1}\pa^{\mu}A = 0, \\
\pa^{\mu}(\rho G^{-1}\pa_{\mu}G)
	+ \rho G^{-1}\pa_{\mu}A \pa^{\mu}A^T \nonumber \\
+ \rho G^{-1} (\pa_{\mu}B - {1\over 2}\pa_{\mu}A A^T
	+ {1\over 2}A \pa_{\mu}A^T)G^{-1}(\pa^{\mu}B
	-{1\over 2}\pa^{\mu}AA^T + {1\over 2}A\pa^{\mu}A^T) =0.
\label{feqn3}
\end{eqnarray}
The equations (\ref{feqn1})-(\ref{feqn3})
are the matrix generalizations of the eqns.(7.7)-(7.10) of
\cite{kinn1} with the identifications $A_1 =B_1 = {A\over 2}$
and $\psi_{11} + 2 A_1 B_1 \equiv B$.
We have thus shown a close relationship between the equations of motion
of the heterotic string theory and the equations of the dual
fields in the Einstein-Maxwell theory.
These identifications allows us to apply the whole formulation
of Kinnersley et al to the present problem with some
variations. For example, unlike in \cite{kinn1,kinn2},
in our case, the analogue of the
component $A_2$ of the gauge field, namely $\tilde{A}$,
does not represent a new degree of freedom.

We can now define two potentials $\psi$ and $\tilde{A}$:
\bea
\label{si7}
G^{-1} (\dmu B + \dmu C - \dmu A A^T ) G^{-1} & = & \rho^{-1}
\epsilon_{\mu \nu} \pa^\nu \psi, \\
\label{si5}
G^{-1} \dmu A - G^{-1} (\dmu B + \dmu C - \dmu A
A^T ) G^{-1} A & =& \rho^{-1} \epsilon_{\mu \nu} \pa^\nu (
\tilde{A} - \psi A),
\eea
such that equations (\ref{feqn1}) and (\ref{feqn2}) are the
integrability conditions for $\psi $ and $\tilde{A}$. One can
also define another independent
potential from equation (\ref{eqn4}), but we
introduce it in the later sections.

Inserting $\tilde{A}$ and $\psi$ into the equations of motion
one can rewrite them as divergence equations containing the
dual potentials. By using
(\ref{si5}) in itself once, one can write the following
equations:
\bea
\label{dual1}
\epsilon_{\mu \nu} \pa^\nu \tilde{A} & = & \left( \rho G^{-1} +
\psi \frac{G}{\rho} \psi \right) \pa_\mu A - \psi \frac{G}{\rho}
\pa_\mu \tilde{A}, \\ \label{dual2}
\epsilon_{\mu \nu} \pa^\nu A & =& -\frac{G}{\rho}
\left(\pa_\mu \tilde{A} - \psi \pa_\mu A \right).
\eea
One can now use $G$, $\psi$ and
$\rho$ to rewrite the equations (\ref{feqn1})--(\ref{feqn3}).
In this way one obtains a set of new equations:
\begin{eqnarray}
	\label{deqn1}
	\pa^{\mu}[\frac{G}{\rho} (\pa_{\mu} \tilde{A} -
	\psi \pa_{\mu} A)] = 0,	\\
	\label{deqn2}
	\pa^{\mu}[\rho G^{-1}\pa_{\mu} A
	- \psi \frac{G}{\rho}(\pa_{\mu}\tilde{A}
	- \psi \pa_{\mu}A)] = 0,	\\
	\label{deqn3}
	\pa^{\mu}[\frac{G}{\rho} \pa_{\mu}\psi G
	+ \frac{G}{\rho}(\pa_{\mu}\tilde{A} - \psi \pa_{\mu}A)A^T]
	= 0,	\\
	\label{deqn4}
	\pa^{\mu}(\rho G^{-1}\pa_{\mu}G)
	- \pa^{\mu}\psi \frac{G}{\rho}\pa_{\mu}\psi G
	+ \rho G^{-1} \pa_{\mu}A \pa^{\mu}A^T = 0.
\end{eqnarray}
These are once again of the same form as the equations
in the Einstein-Maxwell theory.
For example, the form of our eqns.(\ref{deqn1})-(\ref{deqn4})
match with those of (7.1)-(7.4)
of \cite{kinn1} with the
identifications $A_1 \equiv {A\over 2}$,
$A_2 \equiv {\tilde{A}\over 2}$ and
$\omega \equiv \psi$.
Other equations, namely the analogues of (7.5) and (7.6)
in \cite{kinn1}, are derivable from the above ones and have
the same form.

At this point we note a major difference between the situations in
the Einstein-Maxwell theory and string case. We find that the
roles of the original and dual variables as compared to
\cite{kinn1,kinn2,kinn3} have changed in our case. Namely the
$B$ field of string theory plays the role of the dual
variable $\psi_{11}$ in \cite{kinn1} and our $\psi$ plays the
role of the twist potential $\omega$ parameterizing the metric in
\cite{kinn1}. As a result the Ehlers and the Harrison
transformations as written in \cite{kinn1} can not be directly
copied to our case, since they will simply correspond to our
T-duality transformations in equations (\ref{o824}).
Our task, on the
other hand,  is to find the symmetry transformations
with non-local action
on $B$.

One can also
combine the two equations (\ref{dual1}) and (\ref{dual2})
into a single one:
\be \label{dmphi}
\dmu \Phi = - M L \epsilon_{\mu \nu} \pa^\nu \Phi,
\ee
where $M$ is a $ 2d \times 2d$ matrix
\begin{equation}
M =  \left(
\begin{array}{cc}
-\rho G^{-1} - \psi \frac{G}{\rho} \psi &  \psi \frac{G}{\rho} \\
-\frac{G}{\rho} \psi  &  \frac{G}{\rho}
\end{array} \right),
\end{equation}
with $ L M L = - M^{-1} $ and $\Phi$ is an $n \times n$ matrix
written as
\be
\label{defphi}
\Phi = \left( \begin{array}{c} \tilde{A} \\ A  \end{array} \right).
\ee
Equation (\ref{dmphi}) also implies
\be
	\label{phieqn}
	\pa^{\mu}(M^{-1}\pa_{\mu}\Phi) = 0,
\ee
and is equivalent to the two equations (\ref{deqn1})
and (\ref{deqn2}).
Similarly the two equations (\ref{deqn3}) and (\ref{deqn4}) can be
combined as \be \label{correct1}
\pa^\mu \left[ \rho M^{-1} \dmu M - M^{-1} \dmu \Phi \Phi^T
\right] = 0.
\ee
Thus, starting from our original set of equations of motion, we
have arrived at a formulation of the equations that consists in
the variables $G$, $\psi$, $A$, and $\tilde{A}$.
Unlike $B$, $A$ has not been eliminated in favour of its dual in
equations of motion. Setting
the gauge fields to zero we also retrieve the expression for $M$
as well as the equations of motion
of the dual fields given in \cite{ray}.

It can be shown\cite{prep} that the dualized equations of motion
(\ref{deqn1})-(\ref{deqn4}) naturally follow from
the dimensional compactification of the $N=1$ supergravity-
super-Yang-Mills action in ten dimensions
\cite{sch,boonstra} down to two.
We now study the symmetries of the equations
(\ref{phieqn}), (\ref{correct1}) as well as the
original equations (\ref{eom11})-(\ref{eom22}).
\section{Symmetries of the equations of motion}

We observe that the two equations
(\ref{phieqn})-(\ref{correct1}) are manifestly
invaraint under the following transformations:
\begin{eqnarray}
\label{sym1}
(i) M \rightarrow \Omega M \Omega^T, \quad \Phi \rightarrow
			\Omega \Phi, \quad
	\Omega^T L \Omega = L \\
\label{sym2}
(ii) M \rightarrow M, \quad \Phi \rightarrow \Phi \Omega^T,
\quad	\Omega^T \Omega = I_n \\
\label{sym3}
(iii) M \rightarrow M, \quad \Phi \rightarrow \Phi + constt.
\end{eqnarray}
These transformations are the extensions of the group $G'$
mentioned in \cite{kinn1}. Similarly the $O(8, 24)$ mentioned
in equation (\ref{o824}) is the analogue of the group
$H'$ of \cite{kinn1}. Among $G'$ symmetries,
(i) and (iii)
are the same transformations mentioned in\cite{kinn1}. (ii)
is a new symmetry for us, since $A$ in our case
represents $n$-columns as opposed to
a single function $A_1$ or $A_2$ in \cite{kinn1}.
Together, these are a 496-parameter
global symmetry transformations of the dual
equations of motion for the heterotic strings.
Note that although equations (\ref{phieqn}) and
(\ref{correct1}) are not manifestly invariant under
$O(8, 24)$, we will still be able to
show that these equations have an
affine $\hat{O}(8, 24)$ symmetry and
identify $G'$ as part of this.

In subsection (3.a) below we now give the construction
of the infinite set of conserved currents starting
from the dual equations (\ref{phieqn})-
(\ref{correct1}). The same exercise for
the original field equations (\ref{eom11})-
(\ref{eom22}) is done in section (3.b).

Dealing with the dual equations of motion (\ref{deqn1})-
(\ref{deqn4}) is technically more complicated than the
original equations of motion. This is essentially
due to the fact that  the manifest symmetries in the
dimensional compactification of the dual formulation
of the supergravity-super-Yang-Mills are different
from the original formulation. This has  previously
been observed in four dimensions\cite{sch,boonstra}.
In our case the manifest symmetry is only and
$O(8, 8)$ symmetry. We however show that the full
symmetry in this case is also an affine
$\hat{O}(8, 24)$ symmetry.

\noindent {\bf (3.a) Conserved Currents from the Dual
Equations of Motion}

The basic property needed for generating the infinite
dimensional algebra using the procedure of
\cite{kinn1,kinn2,kinn3} is to
find a set  of variables satisfying the
Ernst equations. The $\Phi$ in equation (\ref{dmphi}) already
satifies this property. To find other quantities which
satisfy similar equations, we introduce a potential $\Psi$ as
\be \label{dmpsi}
 M^{-1} \dmu (\rho M) - M^{-1} \dmu \Phi \Phi^T = - \epsilon_{\mu
\nu} \pa^\nu \Psi,
\ee
which also implies
\be \label{potpsi}
 M (\dmu \Psi + L \dmu \Phi \Phi^T) =  \epsilon_{\mu
\nu} \pa^\nu (\rho M),
\ee
and is equivalent to the equation of motion
(\ref{correct1}). Then
$ \cal H $ defined as ${\cal H} \equiv \rho M + L \Psi$,
satisfies the Ernst equation,
\be \label{linear}
\dmu {\cal H} = - M L \epsilon_{\mu \nu} \pa^\nu {\cal
H}.
\ee

One can now use $\Phi$ and ${\cal H}$ to construct the infinite
hierarchy of potentials in terms of which the infinite
dimensional symmetry algebra will be realized.
Let us note at this point that both $\h$ and $\Phi$ contain
the mixture of
original and dual fields. This mixture of fields is also
crucial for the generation of new solutions. Treating $\cal H$
and $\Phi$ as
our field variables, four bilinear potentials are defined
through the following relations:
\bea \label{potk}
\dmu \k = \Phi^T L \dmu \Phi, \\
\label{potn}\dmu \n = \h^T L \dmu \h, \\
\label{potq}\dmu \q = \Phi^T L \dmu \h, \\ \label{potk4}
\label{potp}\dmu \p = \h^T L \dmu \Phi.
\eea
It can be verified, using (\ref{dmphi}) and (\ref{linear}),
that they satisfy the condition:
\be
\label{condition}
\epsilon_{\mu \nu}\pa^{\nu}\pa^{\mu}{\cal K} =
\epsilon_{\mu \nu}\pa^{\nu}\pa^{\mu}{\cal N}=
\epsilon_{\mu \nu}\pa^{\nu}\pa^{\mu}{\cal Q} =
\epsilon_{\mu \nu}\pa^{\nu}\pa^{\mu}{\cal P} = 0.
\ee
Furthermore one
can define more fields out of the four potentials ${\cal K}$,
${\cal N}$, ${\cal Q}$ and ${\cal P}$ which also satisfy the
Ernst equations. For example the fields $\cal R$ and $\cal U$
defined as
\bea \label{upot}
{\cal U} = \n + \Phi \q + \h L \h, \\ \label{rpot}
{\cal R} = \p + \Phi \k + \h L \Phi,
\eea
satisfy field equations similar to those of $\h$ and $\Phi$,
respectively, viz.
\bea \label{req}
\dmu {\cal R} = - M L \epsilon_{\mu \nu} \pa^\nu {\cal R}, \\
\label{ueq}
\dmu {\cal U} = - M L \epsilon_{\mu \nu} \pa^\nu {\cal U}.
\eea
In deriving the equations (\ref{req}) and (\ref{ueq}) for $\cal
R$ and $\cal U$, we had to use an identity
\be
\label{defz}
\dmu ( L \Psi + \Psi^T L + \Phi \Phi^T ) = - 2 \dmu z L,
\ee
where $\epsilon_{\mu \nu} \pa^\nu z = \dmu \rho $.
$\cal R$ and $\cal U$ can now be used in place of $\Phi$ and $\h
$ to define a second set of potentials as in
(\ref{potk})--(\ref{potk4}) and thence another set
of fields, as in (\ref{upot})--(\ref{rpot}) and so on, each time
increasing the order of fields in
the potentials by one. This bestowes us with an infinite set of
fields $\{ \h^n | n =  1, 2 \cdots \} $ and $\{ \Phi^n | n =
1, 2 \cdots \}$ with
\be
\h^1 = \h   \qquad {\rm and } \qquad \Phi^1 = \Phi,
\ee
each satifying the same field equations:
\bea
\dmu \h^n = - M L \epsilon_{\mu \nu} \pa^\nu \h^n, \\
\dmu \Phi^n = - M L \epsilon_{\mu \nu} \pa^\nu \Phi^n .
\eea
{}From these fields we can once again
define four families of potentials as
[cf. (\ref{potk}) and (\ref{potk4}) ]
\bea
\label{kmn} \dmu \k^{mn}  = \Phi^{Tm} L \dmu \Phi^n, \\
\label{nmn} \dmu \n^{mn} = \h^{Tm} L \dmu \h^n, \\
\label{qmn} \dmu \q^{mn} = \Phi^{Tm} L \dmu \h^n, \\
\label{pmn} \dmu \p^{mn} = \h^{Tm} L \dmu \Phi^n,
\eea
which satisfy equations similar to (\ref{condition}).
These in turn enable us to define the solutions of the field
equations cf. (\ref{dmphi})--(\ref{linear}):
\bea
\h^{n + 1} = \n^{1n} + \Phi \q^{1n} + \h L \h^n, \\
\Phi^{n + 1} = \p^{1n} + \Phi \k^{1n} + \h L \Phi^n,
\eea
as the $ (n + 1) $-th set of fields from the $n$-th set in the
hierarchy. Due to the matrix-valuedness of $A$ in our case,
${\cal K}$ for us is an $n\times n$ matrix  instead of
being a single number as in \cite{kinn1,kinn2}. ${\cal P}$ and
${\cal Q}$ are respectively $2d \times n$ and $n\times 2d$ matrices
and ${\cal N}$ is a $2d\times 2d$ matrix.

Several properties of the potentials follow from their
definitions,
\bea \label{id1}
\k^{mn} + \k^{Tnm} = \Phi^{Tm} L \Phi^n, \\
\q^{mn} + \p^{Tnm} = \Phi^{Tm} L \h^n, \\
\n^{mn} + \n^{Tnm} = \h^{Tm} L \h^n.
\eea
Moreover, the potentials are interrelated by some identities,
namely,
\bea \label{rec1}
\k^{m, n+1} - \k^{m+1, n} = \q^{m1} L \Phi^n + \k^{m1} \k^{1n},
\\
\label{rec2}
\q^{m, n+1} - \q^{m+1, n} = \q^{m1} L \h^n  + \k^{m1} \q^{1n}, \\
\label{rec3}
\p^{m, n+1} - \p^{m+1, n} = \n^{m1} L \Phi^n + \p^{m1} \k^{1n},
\\ \label{rec4}
\n^{m, n+1} - \n^{m+1, n} = \n^{m1} L \h^n + \p^{m1} \q^{1n}.
\eea
Derivation of equations (\ref{id1})--(\ref{rec4}) involve
integration by parts. We can set the constants of integrations
to zero without any loss of generality.
For later convenience, we shall now define the potentials for
all, {\em i.e.}, positive, zero and negative integer, values of
$m$ and $n$.  Let us define
\be
\k^{10} = I_n, \qquad {\rm } \qquad \h^0 = L.
\ee
Then, from the definitions of $\p^{mn}$ and $\n^{mn}$, equations
(\ref{pmn}) and (\ref{nmn}), one can identify
\be
\p^{0n} = \Phi^n \qquad {\rm and} \qquad \n^{0n} = \h^n.
\ee
We shall also define
\be
\k^{1 -p, p} = - \k^{p, 1 - p} = - I_n \qquad {\rm and }
\qquad
\n^{p, - p} = - \n^{- p, p} = L,
\ee
where $ p \geq 1$. All other quantities with $m, n \leq 0 $ are
defined to vanish.
One can now generalize the recursion relations
(\ref{rec1})--(\ref{rec4}) to read, respectively, as:
\bea \label{rec11}
\sum_s \left[ \q^{ms} L \p^{1 - s, n} + \k^{ms} \k^{2 -s, n}
\right] = 0, \\
\sum_s \left[ \q^{ms} L \n^{1 - s, n} + \k^{ms} \q^{2 -s, n}
\right] = 0, \\
\sum_s \left[ \n^{ms} L \p^{1 - s, n} + \p^{ms} \k^{2 -s, n}
\right] = 0, \\ \label{rec44}
\sum_s \left[ \n^{ms} L \n^{1 - s, n} + \p^{ms} \q^{2 -s, n}
\right] = 0,
\eea
where $ s $ runs over all integers, $ - \infty \leq s \leq +
\infty $.

We have thus shown the presence of the four
sets of infinite number of conserved currents in our
two dimensional theory corresponding to the potentials
${\cal K}$, ${\cal P}$,
${\cal Q}$ and ${\cal N}$. In the next section we find the
symmetry transformations corresponding to these infinite
set of conserved currents and the corresponding algebra.
Before that however
we give the same construction of
currents starting with the field equation (\ref{eom11}).

\noindent {\bf (3.b) Currents from the Original Equations
of Motion}

By starting with the original
field equations (\ref{eom11}), one can define a
potential:
\be \label{dmcalsi}
 {\cal M}^{-1} \dmu (\rho {\cal M}) =
- \epsilon_{\mu \nu} \pa^\nu {\hat{\Psi}}.
\ee
Then ${\hat{\cal H}} \equiv \rho {\cal M} + i {\cal L} {\hat{\Psi}}$
also satisfies the Ernst equation:
\be \label{linear2}
\dmu \hat{\cal H} = i {\cal M} {\cal L}
		\epsilon_{\mu \nu} \pa^\nu \hat{\cal H}.
\ee
One can therefore obtain the currents on the same line
as in the previous subsection. We now have
\be
\label{potnn}\dmu {\hat{\cal N}} = {\hat{\cal H}}^T
			{\cal L} \dmu {\hat{\cal H}},
\ee
which is now a $(2d+n)\times (2d+n)$ matrix and
satisfies,
\be
\label{epcaln}
\epsilon_{\mu \nu}\pa^{\nu}\pa^{\mu}{\hat{\cal N}} = 0.
\ee
Similarly by using the identity,
\be
\dmu ( {\cal L} {\hat{\Psi}} + {\hat{\Psi}}^T {\cal L} )
		= - 2 \dmu z {\cal L}.
\ee
one finds that
$
\hat{\cal U} = \hat{\cal N} +
	\hat{\cal H} {\cal L} \hat{\cal H}
$
also satisfies
\be
\dmu \hat{\cal U} = i {\cal M} {\cal L}
	\epsilon_{\mu \nu} \pa^\nu \hat{\cal U}.
\ee
Repeating this process, one gets an infinite set of fields
\{${\hat{\cal H}^n| n = 1, 2 ...}$\} with
$\hat{\cal H}^1 = \hat{\cal H}$
which satisfy the same field equations:
\be
\dmu \hat{\cal H}^n = i {\cal M} {\cal L}
	\epsilon_{\mu \nu} \pa^\nu \hat{\cal H}^n.
\ee
One also has a family of potentials represented by
$(2d+n)\times (2d+n)$ matrices:
\be
 \dmu \hat{\cal N}^{mn} = \hat{\cal H}^{Tm}
	{\cal L} \dmu \hat{\cal H}^n, \\
\ee
which satisfy equations similar to (\ref{epcaln}). This in
turn enables us to define the solutions of the field
equations as:
\be
	\hat{\cal H}^{n+1} = \hat{\cal N}^{1 n} +
			\hat{\cal H} {\cal L}
			\hat{\cal H}^n.
\ee
$\hat{\cal N}^{m n}$ satisfy the identities,
$
\label{idcaln}
\hat{\cal N}^{m n} + \hat{\cal N}^{Tn m} = \hat{\cal H}^{Tm} {\cal L}
			\hat{\cal H}^n
$
and are interrelated by
\be
\label{rechh}
\hat{\cal N}^{m, n+1} - \hat{\cal N}^{m+1, n} =
	\hat{\cal N}^{m1} {\cal L} \hat{\cal H}^n.
\ee
Once again potential for all, {\it i.e.}, positive, zero and
negative integer values of $m$ and $n$ are defined by
setting $\hat{\cal H}^0 = {\cal L}$ which also implies
$\hat{\cal N}^{0 n} = \hat{\cal H}^n$. One also sets
$\hat{\cal N}^{p, -p} = - \hat{\cal N}^{-p, p} = {\cal L}$,
where $p\geq 1$. All other quanitities with
$m,n \leq 0$ are again defined to vanish. One can then
again generalize the recursion relations (\ref{rechh}) to
\be
\sum_s \left[ \hat{\cal N}^{ms} {\cal L} \hat{\cal N}^{1 - s, n}
\right] = 0,
\ee
where $s$ runs over all integers, $-\infty \leq s \leq
+ \infty$.


\section{Symmetry algebra}
In this section we first construct the $\hat{o}(8, 24)$
algebra corresponding to
the currents of section-(3.a) and later on
those corresponding to the currents in section-(3.b).

\noindent {\bf (4.a) Algebra from the Dual Equations}

Let us now note that the equations (\ref{rec11})--(\ref{rec44})
are invariant under three sets of transformations, whose
parameters will be denoted by $\gamma^k$, $\xi^k$ and
$\sigma^k$. These
are respectively $2d\times 2d$, $n \times 2d$ and $n\times n$
matrices and
are the analogues of $\gamma^k$, $c^k$ and
$\sigma^k$ of \cite{kinn1,kinn2}. It can be checked that the
equations (\ref{rec11})--(\ref{rec44})
are invariant under the
following set of transformations acting on
variables defined in section-(3.a):
\bea
\label{gamma1}
(i) \gamma^{mn}:  \qquad
\k^{mn} & \rt & \k^{mn} - \sum_{i=1}^k \q^{mi} L \gamma^k L P^{k -
i, n}, \\
\q^{mn} & \rt & \q^{mn} + \q^{m, n + k} L \gamma^k - \sum_{i=1}^k \q^{mi}
L \gamma^k L \n^{k - i, n}, \\
\p^{mn} & \rt & \p^{mn} - \gamma^k L \p^{m + k, n} - \sum_{i=1}^k \n^{mi}
L \gamma^k L \p^{k - i, n}, \\
\n^{mn} & \rt & \n^{mn} - \gamma^k L \n^{m + k, n} + \n^{m, n + k} L
\gamma^k - \sum_{i=1}^k \n^{mi} L \gamma^k L \n^{k - i, n},
\label{gamma4}
\eea
where $\gamma^k $ is a $2d \times 2d $ antisymmetric matrix.
\bea \label{xi1}
(ii) \xi^k :  \qquad
\k^{mn} & \rt& \k^{mn} - \xi^k L \p^{m + k - 1, n} - \q^{m, n
+ k - 1} L \xi^{Tk} - \sum_{i=1}^k \k^{mi} \xi^k L \p^{k - i, n}
\nonumber \\ & + & \sum_{i=1}^{k-1} \q^{mi} L \xi^{Tk} \k^{k - i, n}, \\
	\label{xi2}
\q^{mn} & \rt & \q^{mn} - \xi^k L \n^{m + k - 1, n} + \k^{m,
n + k} \xi^k - \sum_{i=1}^k \k^{mi} \xi^k L \n^{k - i, n}
\nonumber \\ & + & \sum_{i=1}^{k-1} \q^{mi} L \xi^{Tk} \q^{k - i, n},  \\
\label{xi3}
\p^{mn} & \rt & \p^{mn} + \xi^{Tk} \k^{m + k, n} - \n^{m, n + k - 1}
L \xi^{Tk} - \sum_{i=1}^k \p^{mi} \xi^k L \p^{k - i, n} \nonumber
\\ & + & \sum_{i=1}^{k-1} \n^{mi} L \xi^{Tk} \k^{k - i, n}, \\
\label{xi4}
\n^{mn} & \rt & \n^{mn} + \p^{m, n + k} \xi^k + \xi^{Tk} \q^{m +
k, n} - \sum_{i=1}^k \p^{mi} \xi^k L \n^{k - i, n} \nonumber
\\ & + & \sum_{i=1}^{k-1} \n^{mi} L \xi^{Tk} \q^{k - i, n},
\eea
where $\xi^k$ is a $2d \times n $ real matrix.
\bea
\label{sigma}
(iii) \sigma^k : \qquad
\k^{mn} & \rt & \k^{mn} - \sigma^k \k^{m + k, n} + \k^{m, n + k}
\sigma^k  -  \sum_{i=1}^k \k^{mi} \sigma^k \k^{ k - i + 1, n}, \\
\q^{mn} &\rt& \q^{mn} - \sigma^k \q^{m + k, n} - \sum_{i=1}^k \k^{mi}
\sigma^k \q^{ k - i + 1, n}, \\
\p^{mn} &\rt& \p^{mn} + \p^{m, n + k } \sigma^k - \sum_{i=1}^k \p^{mi}
\sigma^k \k^{k - i + 1, n}, \\
\label{sigma4}
\n^{mn} & \rt & \n^{mn} - \sum_{i=1}^k \p^{mi} \sigma^k \q^{ k - i +
1, n},
\eea
where $\sigma^k $ is an $n \times n $ antisymmetric matrix.

It can be verified that the transformations
(\ref{gamma1})-(\ref{sigma4}) keep the recursion relations
(\ref{rec11})-(\ref{rec44})
as well as the Ernst equations invariant and therefore
generate solutions of (\ref{dmphi}) and (\ref{linear})
through these recursions.
To show the invariance of the Ernst equations, one needs
to integrate the identity (\ref{defz}) with vanishing
integration constant.

The symmetry algebra can be obtained by using the transformation
rules given above through their commutators. We get
after some algebra:
\bea
\label{alg1}
\left[ \delta \sigma^k , \delta \sigma^l \right] =\delta
\sigma^{k + l} &,& \qquad
\sigma^{k + l} =  ( \sigma^k \sigma^l - \sigma^l \sigma^k ) \\
\left[\delta \sigma^k , \delta \gamma^l \right] =
0 &, & \\
\left[\delta \sigma^k , \delta \xi^l \right] =\delta \xi^{ k +
l} &, & \qquad
\xi^{k + l} =  \sigma^k \xi^l \\
\left[\delta \gamma^k , \delta \gamma^l \right] =\delta \gamma^{
k + l } &, &
\gamma^{k + l} =  ( \gamma^k L \gamma^l - \gamma^l L \gamma^k ) \\
\left[ \delta \xi^k , \delta \gamma^l \right] = \delta \xi^{k +
l } & , &  \xi^{ k + l } = \xi^k L \gamma^l \\
\left[\delta \xi^k , \delta \xi^l \right] = \delta \gamma^{ k + l
- 1} + \delta \sigma^{ k + l - 1} &,& \gamma^{k + l - 1} = (
\xi^{Tk} \xi^l - \xi^{Tl} \xi^k )
 \nonumber \\ &{\rm and} & \sigma^{k + l - 1} = ( \xi^k
L \xi^{Tl} - \xi^l L \xi^{Tk} ). \label{alg6}
\eea
One can also write this symmetry algebra in a
compact form by introducing a set of
$2d +n$-dimensional antisymmetric matrices generating
the symmetry transformations:
\begin{equation}
\label{gammaa}
\Gamma^k_1 = \left(\begin{array}{cc} \gamma^k & 0 \\
                          0 & 0 \\
\end{array}\right)  \quad
\Gamma^k_2 = \left(\begin{array}{cc} 0 & 0 \\
                          0 & \sigma^k \\
\end{array}\right) \quad
\Gamma^k_3 = \left(\begin{array}{cc}  0 & {\xi^k}^T \\
                                    -{\xi^k} & 0 \\
\end{array}\right)
\end{equation}
It can then be shown that the algebra in equations
(\ref{alg1})-(\ref{alg6}) combine into
\be
\label{alga}
[\delta \Gamma^k, \delta \Gamma^l]
= \delta \Gamma^{k+l},
\ee
where $\Gamma^{k+l} = \Gamma^k {\cal L} \Gamma^l
- \Gamma^l {\cal L} \Gamma^k$.

One can identify
the $\Gamma^k_a$'s in eqn.(\ref{gammaa}) as the generators of the
$O(8, 24)$ transformations and therefore
equation (\ref{alga}) represents an
affine $\hat{o}(8, 24)$ symmetry algebra.

\noindent{\bf (4.b) Algebra from the Original Equations}

The symmetry transformations of the infinite set of
potentials $\hat{\cal N}^{mn}$ as constructed
in section (3.b) from the
field equation (\ref{eom11}) is given by
\be
\hat{\cal N}^{mn}  \rt  \hat{\cal N}^{mn} -
\hat{\Gamma}^k {\cal L} \hat{\cal N}^{m + k, n}
+ \hat{\cal N}^{m, n + k} {\cal L}
\hat{\Gamma}^k - \sum_{i=1}^k \hat{\cal N}^{mi}
{\cal L} \hat{\Gamma}^k {\cal L} \hat{\cal N}^{k - i, n},
\ee
where $\hat{\Gamma}^k$ is a $(2d+n)\times (2d+n)$
antisymmetric matrix. Once again we have verified that
these transformations leave the recursion relations, as well as
the field equations, invariant.
In this case one directly gets the
$\hat{o}(8, 24)$ symmetry algebra:
\be
\label{alga2}
[\delta \hat{\Gamma}^k, \delta \hat{\Gamma}^l]
= \delta \hat{\Gamma}^{k+l},
\ee
where $\hat{\Gamma}^{k+l} = \hat{\Gamma}^k {\cal L}
\hat{\Gamma}^l - \hat{\Gamma}^l {\cal L} \hat{\Gamma}^k$.

\section{Discussions}

In this paper we have shown the connection of the
two dimensional heterotic
string equations of motion and its symmetries with those of
the axisymmetrix Einstein-Maxwell equations. The
existence of an infinite
number of conserved currents and $\hat{O}(8, 24)$ symmetry
in the two dimensional heterotic string theory
has also been shown by using this analogy.
The construction of the infinite symmetry structure
was given for both
the original equations of motion of the
heterotic string as well as the one written after dualization
and the elimination of the field $B$.
The construction of the $\hat{o}(8, 24)$
for the dual equations of motion was possible, even though
they are not manifestly invariant under an $O(8, 24)$ symmetry
group. As already mentioned, these dual equations of motion
have a natural interpretation in terms of the compactification
of the dual $N=1$ supergravity-super Yang-Mills action
in ten dimensions.
Our results therefore imply a Geroch symmetry in the dual
formulation as well. As expected, due to the absence of
a manifest $O(8, 24)$ symmetry in the equations of motion,
the algebraic complications for this case in sections-
(3.a) and (4.a) turned out to be considerably more
than for the original equations of motion.

In the present procedure, the action
of the symmetry transformations
on our original fields in
equation (\ref{defcalm}) can also be found.
This can be done for the symmetry transformations in
section-(4.a) by identifying $\Psi_{11}$ as $B$ through
the dualization relations. Other fields, $G$ and $A$
already appear in the transformation laws through
$\Phi$ and $M$. For example,
The transforamtion by the parameters $\gamma^0$ is simply the
infinitesimal version of the $O(d, d)$ transformations
in equation (\ref{sym1}).
This can
be seen by writing down its action  on the potentials.
They can be shown to transform as
\be
{\cal N} \rightarrow \Omega {\cal N} \Omega^T,
\quad
{\cal P} \rightarrow \Omega {\cal P}, \quad
{\cal Q} \rightarrow  {\cal Q} \Omega^T, \quad {\rm and}
\quad {\cal K} \rightarrow {\cal K}.
\ee
The infitesimal form of these transformations
matches with (\ref{gamma1})-(\ref{gamma4}),
since all the
quadratic terms in these equations vanish for $k=0$.
Similarly it can be shown that the
transformations (\ref{sym2})
correspond to $\sigma^0$ and (\ref{sym3}) to $c^0$.

The nontrivial action of $\gamma^0$
is given by the Ehlers transformation
whch is represented by $\Omega$ of the form
$\Omega_2$ in equation (\ref{omega}).
The situation here is different from
\cite{kinn1,kinn2} where the Ehlers transformation
is represented by $\gamma^1$. This is because,
unlike in \cite{kinn1,kinn2} our
$M$ already represents the dual variables and a linear
transformation like $\gamma^0$ induces non-local
transformation of $B$. The action of these transformations
on $B$ can also be found out explicitly, but we do
not give them here. For vanishing $A, \tilde{A}$ backgrounds,
the action of $\gamma^0$, or
equivalently the transformations of equation
(\ref{sym1}) reduces to the Ehlers transformation
in \cite{ray}.

On the other hand, if one proceeds by using the results
of our sections-(3.b) and (4.b), namely the infinite
extensions of the original equations of motion, then the
Ehlers as well as the Harrison transformations are identified
as the $\hat{\Gamma^1}$ transformations. In this case
$\hat{\Gamma}^0$ is simply the T-duality rotations of
equation (\ref{o824}).

We now discuss the connection between the two sets of
conserved currents as well as symmetry structure shown in
sections (3.a)-(4.a) and (3.b)-(4.b). For the later
one, the $O(8, 24)$ symmetry is manifest, whereas for the
former ones this has been shown only after studying the
algebraic structure. It may be possible in the former
case also to combine the
fields appearing in the dual equations of motion into a
manifestly $O(8, 24)$ invariant form. However it probably
requires the elimination of one of the degree of freedom
out of $A$ and $\tilde{A}$. This can be done
possibly by
finding a gauge symmetry and then by making a proper
choice of gauge. It may be interesting to study this
aspect further.

In this context,
it is also important to find out the connection
between the dual variables ${\hat{\Psi}}$, which appears
as the dual of ${\cal M}$ and the matrices
$M$ and $\Phi$ which appear in the dual set
of equations of motion. By comparing equations
(\ref{si7}), (\ref{si5}) with the components of
(\ref{dmcalsi}) one recoginizes ${\hat{\Psi}}_{21}
\equiv \psi$ and ${\hat{\Psi}}_{23} \equiv
(\tilde{A} - \psi A)$. However
a complete identification of the
components of ${\hat{\Psi}}$ with those of
$M$ and $\Phi$ involves potentials other than
the two mentioned above. Similar problem also occurs
in the identification of $\Psi$ in terms of the
fileds $G$, $B$ and $A$.

It has been pointed out in \cite{kinn1} that
the infinite dimensional symmetry
$K'$ identified as $\hat{s}l(3, R)$, which is the
analogue of $\hat{O}(8, 24)$ in our case, can be
obtained by the combined operations of
finite dimensional symmetries, $G'$ and $H'$.
Extensions to infinite dimensional symmetry by the
combined operations of T- and S-duality has also
been noted recently\cite{bak2}.
It will therefore be interesting to identify
$G'$ as a combination of T- and S-duality.
In this context
we have already identified $H'$ as the T-duality
symmetry of the two dimensional heterotic string.

There are many other interesting avenues to explore in
this area. First, one should be able to write down the
finite action of all the transformations
$\gamma^k, \xi^k, \sigma^k$ and $\hat{\Gamma}^k$
rather explicitly by following the later works of
Kinnersley et al. In particular the exponentiation of
$\hat{\Gamma}^1$ will give the finite form of the
Ehlers and Harrison transformations.
This will be important for generating the
solutions of the heterotic strings similar to the
Einstein-Maxwell case\cite{ste}.

In the light of recent progress made in understanding
the duality symmetries of the $N=2$ supersymmetric
theories,
it should also be of interest to study the
symmetries of other two dimensional string theories
such as the ones compactified on manifolds other than
the tori, such as $K_3 \times T^4$ or a product of
Calabi-Yau with $T^2$ and so on. Progress on these
topics will hopefully be reported in near future.

\end{document}